\documentstyle[11pt]{article}  
\input{epsf}

\def\spacing #1{\small \renewcommand{\baselinestretch}{#1} \normalsize}

\newcounter{nr}
\newcounter{subnr}[nr]
\newcounter{subsubnr}[subnr]
\newcommand{\nsection}[1]{\stepcounter{nr} \par \vspace{2ex}
  \begin{center} {\thenr.~~#1} \end{center}
  \nopagebreak \vspace*{-1.3ex}}
\newcommand{\nsectionnon}[1]{\par \vspace{2ex}
  \begin{center} {#1} \end{center}
  \nopagebreak \vspace*{-1.3ex}}

\def\thebibliography#1{\par \vspace{2ex}
 \begin{center} {REFERENCES} \end{center}
 \nopagebreak \vspace*{-1.3ex} \list
 {\arabic{enumi}.}{\settowidth\labelwidth{3ex}\leftmargin\labelwidth
 \advance\leftmargin\labelsep
 \usecounter{enumi}}
 \def\newblock{\hskip .11em plus .33em minus -.07em}
 \sloppy
 \sfcode`\.=1000\relax}

\begin{document}
\spacing{0.9091}
\parskip=0ex \parindent=2ex
\small

\twocolumn[{%
\begin{center}
    {\large \bf  HELIOSEISMIC DETERMINATION OF OPACITY CORRECTIONS}\\[4ex]   
    {\normalsize \bf
S.C. Tripathy$^1$, Sarbani Basu$^2$ and J. Christensen-Dalsgaard$^{2,3}$}\\[4ex]
    \begin{minipage}[t]{16cm}
      \begin{tabbing}               
        $^1$Udaipur Solar Observatory, Physical Research
Laboratory, PO Box No. 198, Udaipur 313 001, India\\
$^2$Teoretisk Astrofysik Center, Danmarks Grundforskningsfond,
DK-8000 Aarhus C, Denmark\\
$^3$Institut for Fysik og Astronomi, Aarhus Universitet,
DK-8000, Aarhus, Denmark
      \end{tabbing}
    \end{minipage}
\end{center}}]



\nsection{INTRODUCTION}
Accurate measurements of oscillation frequencies of the Sun are 
providing detailed information about the structure of the Sun.
These frequencies can be used to test the basic input physics, like
opacities, equation of state and nuclear energy generation rates which
are required to construct theoretical solar models.
Here we concentrate on the effects of opacity.
Throughout we consider models calibrated to solar radius and luminosity,
by adjusting the initial composition and the mixing-length parameter.

\vskip -10 true pt
\nsection{METHOD}
We consider opacity modifications given by
\begin{equation}
\log \kappa = \log \kappa_0 + f(T, T_0) \; , 
\end{equation}
where $\kappa_0$ is the unmodified opacity,
obtained as a function of $T$, density $\rho$ and composition
and the function $f(T,T_0)$ has the form
\begin{equation}
f(T,T_0) = A \exp \left[-\left({\log T - \log T_0\over \Delta}\right)^2\right]
\; ,
\end{equation}
where $\log$ is logarithm to base 10. 
The constants $A$ and $\Delta$ set the magnitude and width of the opacity
modification.
By varying $T_0$, we can investigate the effects of 
changes in different parts of the models. 

Assuming a linear relation between small
modifications 
in the opacity and the response of the model 
(see Tripathy \& Christensen-Dalsgaard 1995),
the effect of an opacity change
$\delta \log \kappa(T)$ 
for any quantity $F$ can be approximated by the relation
\begin{equation}
{\delta F \over F} 
= \int K_F(T) \delta \log \kappa(T) {\rm d} \log T \; .
\label{fker}
\end{equation}
The kernel $K_F(T)$ may be 
estimated from the change $\delta F$ 
corresponding to the opacity modification given 
in equation (1) for sufficiently small $A$ and $\Delta$.

For this work we consider $F=c(r)^2$, the squared sound speed;
thus $\delta c^2/c^2$ is the relative squared sound-speed  difference
between the Sun or a test model and the reference model.
This quantity can be obtained by a straightforward inversion of 
solar oscillation frequencies (e.g., Basu {\it et al.}, 1996).
It can be related to an intrinsic opacity difference $\delta \log \kappa$
through equation~(\ref{fker}), with a kernel $K_c(r, T)$ 
which may be estimated by applying modifications of the form
given in equation~(2) for a range of $T_0$.
If the sound-speed difference between the Sun and the model
is assumed to arise solely from opacity errors, these may
then be determined by inversion of the relation between
$\delta \log \kappa$ and $\delta c^2/c^2$.

Here we expand the opacity correction in terms of 
the basis functions given by equation~(2), as
\begin{equation}
\delta \log \kappa(T) = \sum_{i=1}^N b_i f(T, T_i) \; . 
\end{equation}
The resulting change in the squared sound speed can be written as
\begin{equation}
\left({\delta c^2 \over c^2} \right) (r)
=\sum_{i=1}^N b_i \psi_i(r) \; ,
\label{deltac}
\end{equation}
where $\psi_i(r)$ is the change $(\delta c^2 / c^2)(r)$ resulting from
applying the opacity change $\delta \log \kappa = f(T, T_i)$.
It was determined by carrying out a solar evolution calculation
with the modified opacity.
Having thus obtained 
$\psi_i(r)$ for a suitable set of $T_i$,
equation~(\ref{deltac}) may be fitted to the actual $\delta c^2/c^2$
by means of a regularized least-squares fit
to determine the coefficients $b_i$, and hence the opacity
correction $\delta \log \kappa$.
We have calculated $\psi_i(r)$ at 57 values of $T_i$,
with $\log T_i=6.2, \ldots, 7.19$, $\Delta=0.02$ and $A = 0.02$.

The procedure was tested by applying it to differences between two models:
the reference model was computed with the Cox \& Tabor (1976)
opacities, while the test model used the Los Alamos Opacity Library
(Huebner {\it et al.} 1977).
Frequency differences between the two models, for the mode set used in the
analysis of solar data,
were inverted to infer the sound-speed differences,
and the result was fitted to obtain the corresponding opacity change.
The resulting inferred $\delta \log \kappa$ was in good agreement with
the actual intrinsic opacity difference between the two tables
(at fixed $T$, density $\rho$ and composition).
Thus we conclude that the method allows a determination of
the opacity correction required to match a given sound-speed difference.

\begin{figure}[tbp]
\vskip -0.2cm
\hskip 0.5cm
\epsfxsize=6.5 true cm\epsfbox{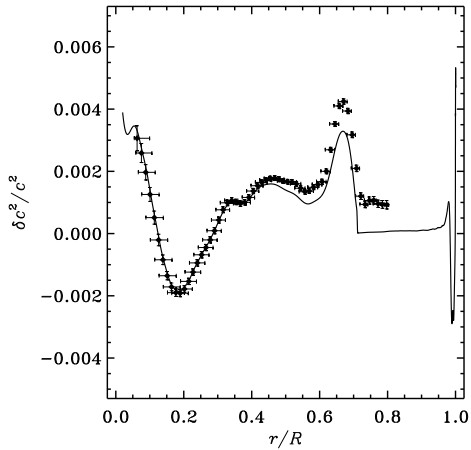}\vskip -0.5 true cm
\caption{
Comparison between the inferred difference in squared sound
speed between the Sun and the model (Model S of
Christensen-Dalsgaard {\it et al.} 1996), shown with symbols,
and the model change resulting from an evolution calculation
where the opacity change in Fig.~2 was added to the opacity tables.
The differences are shown in the sense (Sun) -- (reference model).} 
\vskip -10 true pt
\end{figure}

\vskip -10 true pt
\nsection{RESULTS FOR SOLAR DATA}
We have applied the procedure to sound-speed differences
between the Sun and a solar model.
The data, model and inversion results are described 
by Basu {\it et al.}, (1996).
The data are a combination of low-degree BiSON data
and LOWL data of degrees between 3 and 99.
The reference model (Model S of Christensen-Dalsgaard {\it et al.}\ 1996)
used OPAL opacities (e.g. Iglesias {\it et al.} 1992), and included
settling of helium and heavy elements.
The inversion was carried out by means of the Subtractive
Optimally Localized Averaging technique.
The inferred differences in squared sound speed are shown in Figure 1.

The opacity modification resulting from a fit to the sound-speed difference
is shown in Fig.~2.
This indicates that a sharp feature in $\delta \log \kappa$ is required
just beneath the convection zone, with a second peak in the core.
To test that this opacity modification does in fact reproduce
the inferred sound-speed difference, we have subsequently
recomputed the solar evolution model, adding $\delta \log \kappa(T)$
to the opacity as interpolated from the tables.
In Fig.~1, the solid line shows the resulting difference
in squared sound speed, relative to the reference model.
It is evident that this is in fact quite close to the sound-speed
differences inferred from the inversion.

\begin{figure}[tbp]
\vskip -0.2cm
\hskip 0.5cm
\epsfxsize=6.5 true cm\epsfbox{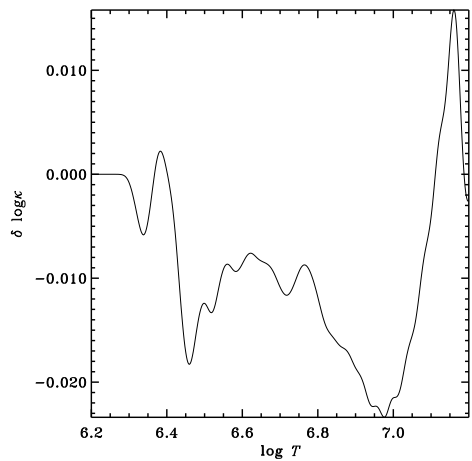}\vskip -0.5 true cm
\caption{
Opacity difference fitted from inferred sound-speed
difference between the Sun and Model S of 
Christensen-Dalsgaard {\it et al.} (1996).}			
\vskip -10 true pt
\end{figure}

\vskip -10 true pt
\nsection{DISCUSSION}
We have attempted to infer the corrections to the opacity
required to match the sound speed in a state-of-the-art solar
model to that inferred from helioseismic inversion. We find that
the sound-speed difference  can to a large extent be reproduced
by a change in the opacity in the radiative interior.
The required change, of order of a few percent, is probably within
the general level of uncertainty of current opacity calculations.
Since the energy transport in the 
solar convection zone is independent of opacity, the sound-speed
difference does not give us any information about the opacity in that region. 

It must be noted that the sound-speed difference between two models
is essentially insensitive to a change in the opacity
by a constant factor within the radiative interior.
The change in
opacity is largely compensated by a change in the composition, 
the sound-speed profile remaining approximately the same.
Thus in the fit in Fig.~2,
$\delta \log \kappa$ is determined only to within a constant.
The freedom in the level of the relative opacity change could, for example,
be utilized to choose a solution for which the change in the
initial hydrogen abundance is small.  

We also note that rather similar effects on the sound speed can be
obtained through suitable weak mixing of the solar interior or
at the base of the convection zone (Basu \& Antia, 1994).
Therefore, it cannot be claimed that
the opacity errors are the only, or even
the dominant, actual error in current solar-model calculations. 
A more detailed
investigation of the individual contributions to the opacity,
and their likely uncertainty, is required to test whether the
specific shape of $\delta \log \kappa$ is realistic.

\vskip -10 true pt
\footnotesize
\nsectionnon{ACKNOWLEDGEMENTS}
This work was
supported by the Danish National Research Foundation through its establishment
of the Theoretical Astrophysics Center.
SCT acknowledges financial support from the Department of
Science and Technology, Government of India.

\end{document}